\documentclass[11pt]{article}
\usepackage[margin=1in]{geometry}
\usepackage{amsmath,amssymb,amsthm,mathtools}
\usepackage{booktabs}
\usepackage{microtype}
\usepackage{hyperref}
\usepackage{enumitem}
\usepackage[T1]{fontenc}
\usepackage{lmodern}
\usepackage{bm}
\usepackage{array}
\usepackage{longtable}

\hypersetup{colorlinks=true,linkcolor=blue,citecolor=blue,urlcolor=blue}
\newtheorem{theorem}{Theorem}[section]

\newcommand{\Hcal}{\mathcal H}

\newcommand{\Zcal}{\mathcal Z}
\newcommand{\Acal}{\mathcal A}
\newcommand{\Dcal}{\mathcal D}

\title{Path Integrals on Functional Lattices}
\author{Alexey A. Kryukov\\University of Wisconsin--Milwaukee}
\date{}

\begin{document}
\maketitle

\begin{abstract}
Path integrals in quantum field theory are commonly approached through sequences of finite-dimensional approximations. The existence and regulator independence of the corresponding infinite-dimensional limit, however, do not follow from the finite-dimensional construction itself and require separate analysis. We introduce an approximation scheme in which spacetime remains continuous and the field is expanded in a discrete complete family of localized functions, referred to as a functional lattice. Truncation to finite-dimensional functional subspaces turns the path integral into an ordinary finite-dimensional integral while leaving the continuum differential operator unchanged.
For the massive free Euclidean scalar field, we prove convergence of the finite-dimensional covariance operators and normalized generating functionals by a Galerkin argument. The limit is independent of the particular compatible functional lattice: any sequence of finite-dimensional subspaces dense in the quadratic-form domain converges to the same continuum covariance and generating functional. A corresponding direct Minkowski construction is obtained at fixed Feynman regulator $i\epsilon$, followed by the distributional limit $\epsilon\downarrow0$. For interacting theories, each finite truncation gives an ordinary non-Gaussian integral. At fixed perturbative order, convergence follows when the finite-resolution graph integrands converge pointwise and admit an integrable majorant independent of the functional resolution. Numerical calculations for one-dimensional $\phi^4$ theory approach the known first- and second-order mass-gap coefficients, while four-dimensional calculations exhibit the expected mass divergence and are consistent with the standard logarithmic wave-function renormalization.
The construction therefore provides a continuum-space approach to functional integration in which convergence can be established directly in the cases considered here. In the free theory, the continuum limit is also independent of the particular compatible functional lattice.
\end{abstract}

\section{Introduction}

The path-integral formulation is indispensable in quantum field theory \cite{GlimmJaffe,PeskinSchroeder,ZinnJustin}. Formally, for a Euclidean scalar field $\phi$ one writes
\begin{equation}
Z[J]=\int \mathcal D\phi\,\exp\left[-S[\phi]+(J,\phi)\right],
\end{equation}
where $S[\phi]$ is the action functional, $J$ is the source, and the integration is assumed to be over all possible field configurations.
In applications this expression is normally replaced by finite-dimensional approximations, obtained for example by discretizing spacetime or imposing a momentum cutoff \cite{GlimmJaffe,PeskinSchroeder,ZinnJustin}. The passage from finite-dimensional integrals to an infinite-dimensional theory is, however, a separate mathematical problem. Finite dimensionality alone does not establish existence, uniqueness, or regulator independence of the continuum limit. Rigorous continuum constructions are known in important special classes of models, particularly within constructive quantum field theory \cite{GlimmJaffe}, but they require estimates specific to the theory under consideration.

Here we consider a different approximation. Rather than discretizing spacetime, we discretize the representation of the field in function space.
Let $\Hcal$ be the one-particle function space and let
\begin{equation}
\{\psi_\alpha\}_{\alpha\in\Lambda}
\end{equation}
be a complete family of localized functions in $\Hcal$, where $\Lambda$ is a discrete index set. We refer to such a discrete complete family as a \emph{functional lattice}. Writing
\begin{equation}
\phi(x)=\sum_{\alpha\in\Lambda}q_\alpha\psi_\alpha(x),
\end{equation}
we truncate to finite subsets $\Lambda_R\subset\Lambda$,
\begin{equation}
V_R=\operatorname{span}\{\psi_\alpha:\alpha\in\Lambda_R\},
\end{equation}
and define
\begin{equation}
\phi_R(x)=\sum_{\alpha\in\Lambda_R}q_\alpha\psi_\alpha(x).
\end{equation}
The finite-resolution functional integral is then simply
\begin{equation}
Z_R[J]=\int_{\mathbb R^{N_R}}\exp\left[-S[\phi_R]+(J,\phi_R)\right]d^{N_R}q,
\qquad N_R=\dim V_R.
\end{equation}
Thus every approximation is an ordinary finite-dimensional integral.

The distinction from a conventional spacetime lattice is important. The coordinate $x$ remains continuous, derivatives remain ordinary continuum derivatives, and the differential operator in the action is not replaced by a finite-difference operator. Finite resolution means only that a finite-dimensional subspace of the full function space is being used.
The principal mathematical point of the present work is not the existence of finite-dimensional approximants; these are standard in path-integral constructions. Rather, for the free field we can prove that the approximants converge, and that the limit is the same for every compatible functional lattice.
For interacting ultraviolet-finite theories, an analogous statement holds graph by graph in perturbation theory. In four-dimensional renormalizable theories, the functional lattice instead provides a continuum-space regulator: regulator-dependent bare quantities diverge as completeness is restored, while the familiar renormalization structure emerges.

Localized Wilson bases are particularly convenient realizations of the construction because they combine orthogonality with simultaneous localization in position and momentum. Wilson orthonormal bases with strong time-frequency localization are well known \cite{DaubechiesJaffardJourne}. The convergence arguments below, however, do not depend specifically on a Wilson basis.

\section{Functional lattices}

Let $\mathcal{H}$ be the one-particle function space and let
\[
V_1\subset V_2\subset\cdots\subset\mathcal{H}
\]
be finite-dimensional subspaces generated by finite subsets of a localized
complete system. Suppose that the quadratic form $a$ is positive definite
on its form domain $\mathcal{V}$, and define the corresponding form norm by
\[
\|u\|_a^2=a(u,u).
\]
We call the sequence $\{V_R\}$ compatible with $a$ if
\[
\bigcup_{R=1}^{\infty}V_R
\]
is dense in $\mathcal{V}$ with respect to the norm $\|\cdot\|_a$.

For the massive Euclidean scalar field,
\[
K=-\Delta+m^2,\qquad m>0,
\]
the form domain is
\[
\mathcal{V}=H^1(\mathbb{R}^d),
\]
with
\[
a(u,v)
=
\int_{\mathbb{R}^d}
\left(
\nabla u\cdot\nabla\overline{v}
+
m^2u\overline{v}
\right)\,dx.
\]
In this case $a$ is positive definite, and the form norm is equivalent
to the usual $H^1$-norm. Thus compatibility means that
\[
\bigcup_{R=1}^{\infty}V_R
\]
is dense in $H^1(\mathbb{R}^d)$.

At this stage no basis of $V_R$ is chosen and no orthogonality assumption
is required. The convergence results below depend only on the subspaces
$V_R$ and their density in the form domain. Localization becomes important
for the practical construction of these spaces and for the sparsity or
quasi-locality of the resulting finite-dimensional matrices.

\section{Free Euclidean field}

Let
\begin{equation}
K=-\Delta+m^2,\qquad m>0,
\end{equation}
act in $L^2(\mathbb{R}^d)$ with domain $H^2(\mathbb{R}^d)$.
The operator $K$ is self-adjoint and strictly positive \cite{ReedSimon1,ReedSimon2}:
\begin{equation}
\langle u,Ku\rangle
=
\|\nabla u\|_{L^2}^2
+
m^2\|u\|_{L^2}^2
\ge
m^2\|u\|_{L^2}^2.
\end{equation}
Hence $\ker K=\{0\}$ and $K^{-1}$ exists as a bounded self-adjoint
operator on $L^2(\mathbb{R}^d)$. We denote
\begin{equation}
C=K^{-1}.
\end{equation}

For $J\in L^2(\mathbb{R}^d)$, let
\begin{equation}
u=CJ.
\end{equation}
Equivalently,
\begin{equation}
a(u,v)=\langle J,v\rangle
\qquad
\text{for all }v\in H^1(\mathbb{R}^d).
\end{equation}

The finite-dimensional Galerkin solution $u_R\in V_R$ is defined by
\begin{equation}
a(u_R,v_R)=\langle J,v_R\rangle
\qquad
\text{for all }v_R\in V_R.
\end{equation}
We define the finite-dimensional covariance operator $C_R$ by
\begin{equation}
C_RJ=u_R.
\end{equation}

\begin{theorem}[Convergence of the free covariance]
Suppose that
\begin{equation}
\overline{\bigcup_R V_R}
=
H^1(\mathbb{R}^d),
\end{equation}
where the closure is taken with respect to the form norm $\|\cdot\|_a$.
Then
\begin{equation}
C_RJ\longrightarrow CJ
\end{equation}
in the form norm and hence in $L^2(\mathbb{R}^d)$ for every
$J\in L^2(\mathbb{R}^d)$.
\end{theorem}

\begin{proof}
Galerkin orthogonality gives
\begin{equation}
a(u-u_R,v_R)=0
\qquad
(v_R\in V_R).
\end{equation}
Because $a$ is symmetric and positive definite,
\begin{equation}
\|u-u_R\|_a
\le
\inf_{w_R\in V_R}\|u-w_R\|_a.
\end{equation}
Thus the Galerkin solution is the best approximation to $u$ in $V_R$
with respect to the form norm. Density of $\bigcup_RV_R$ in the form
domain implies
\begin{equation}
\|u-u_R\|_a\longrightarrow0.
\end{equation}
Since
\begin{equation}
m^2\|u-u_R\|_{L^2}^2
\le
\|u-u_R\|_a^2,
\end{equation}
the convergence in $L^2$ follows.
\end{proof}

We now introduce coordinates on $V_R$ in order to write the corresponding
finite-dimensional integral. Let
\begin{equation}
N_R=\dim V_R
\end{equation}
and choose a basis
\[
\{\psi_\alpha\}_{\alpha=1}^{N_R}
\]
of $V_R$. No orthogonality assumption is needed. A field
$\phi_R\in V_R$ can be written uniquely as
\begin{equation}
\phi_R
=
\sum_{\alpha=1}^{N_R}q_\alpha\psi_\alpha,
\end{equation}
where
\[
q=(q_1,\ldots,q_{N_R})^T
\]
is its coordinate vector.

In this basis define
\begin{equation}
(K_R)_{\alpha\beta}
=
a(\psi_\beta,\psi_\alpha)
\end{equation}
and
\begin{equation}
(J_R)_\alpha
=
\langle J,\psi_\alpha\rangle.
\end{equation}
The Galerkin equations then become
\begin{equation}
K_R q=J_R.
\end{equation}
Consequently,
\begin{equation}
C_RJ
=
\sum_{\alpha,\beta=1}^{N_R}
\psi_\alpha
(K_R^{-1})_{\alpha\beta}
(J_R)_\beta,
\end{equation}
and
\begin{equation}
\langle J,C_RJ\rangle
=
J_R^T K_R^{-1}J_R.
\end{equation}

The normalized finite-dimensional Gaussian integral is
\begin{equation}
\mathcal{Z}_R[J]
=
\frac{
\displaystyle
\int_{\mathbb{R}^{N_R}}
\exp\left[
-\frac12q^TK_Rq+J_R^Tq
\right]d^{N_R}q
}{
\displaystyle
\int_{\mathbb{R}^{N_R}}
\exp\left[
-\frac12q^TK_Rq
\right]d^{N_R}q
}.
\end{equation}
Direct Gaussian integration gives
\begin{equation}
\mathcal{Z}_R[J]
=
\exp\left[
\frac12J_R^TK_R^{-1}J_R
\right]
=
\exp\left[
\frac12\langle J,C_RJ\rangle
\right].
\end{equation}

Although the coordinate measure $d^{N_R}q$ depends on the chosen basis,
the normalized integral does not. Indeed, under any invertible linear
change of coordinates in $V_R$, the Jacobian is the same in the numerator
and denominator and therefore cancels. An orthonormal basis is convenient,
and will be used in the numerical calculations below, but it is not
required for the definition of the normalized finite-dimensional
generating functional.

By the convergence theorem,
\begin{equation}
\boxed{
\mathcal{Z}_R[J]
\longrightarrow
\exp\left[
\frac12\langle J,CJ\rangle
\right].
}
\end{equation}
Thus the normalized free functional integral is obtained as the proven
limit of ordinary finite-dimensional integrals.

The normalization is essential here. The determinant factor in the
unnormalized quantity $Z_R[0]$ depends on the choice of finite-dimensional
volume element and is not being claimed to possess a basis-independent
continuum limit. The normalized generating functional and the correlation
functions derived from it are the objects considered here.

\section{Independence of the compatible functional lattice}

The preceding discussion involves two distinct notions of independence.
At fixed $R$, the normalized finite-dimensional generating functional
depends only on the subspace $V_R$ and not on the basis chosen in that
subspace. We now show that, in the continuum limit, the result is also
independent of the particular compatible sequence of subspaces.

Let $\{V_R\}$ and $\{W_R\}$ be two sequences of finite-dimensional spaces
satisfying
\begin{equation}
\overline{\bigcup_RV_R}
=
\overline{\bigcup_RW_R}
=
H^1(\mathbb{R}^d),
\end{equation}
where the closures are taken with respect to the form norm.
Let $C_R^{(V)}$ and $C_R^{(W)}$ denote the corresponding Galerkin
covariance operators.

\begin{theorem}[Functional-lattice independence]
For every $J\in L^2(\mathbb{R}^d)$,
\begin{equation}
C_R^{(V)}J\longrightarrow CJ,
\qquad
C_R^{(W)}J\longrightarrow CJ.
\end{equation}
Consequently,
\begin{equation}
\lim_{R\to\infty}\mathcal{Z}_R^{(V)}[J]
=
\lim_{R\to\infty}\mathcal{Z}_R^{(W)}[J]
=
\exp\left[
\frac12\langle J,CJ\rangle
\right].
\end{equation}
\end{theorem}

\begin{proof}
The convergence theorem applies separately to the two compatible
sequences. Both Galerkin solutions converge to the unique solution of
$Ku=J$, and hence both covariance operators converge to $C=K^{-1}$ on
every $J\in L^2(\mathbb{R}^d)$. The statement for the generating
functionals follows immediately.
\end{proof}

Thus neither the choice of coordinates within a finite-dimensional
subspace nor the choice of a particular compatible functional lattice
affects the limiting free theory. Localized Wilson functions are useful
because of their orthogonality and strong position-frequency localization,
but these properties are computational advantages rather than assumptions
of the convergence and lattice-independence theorems.

\section{Numerical free-field tests}

\subsection{One dimension}

For
\begin{equation}
K=-\frac{d^2}{dx^2}+1,
\end{equation}
the exact covariance kernel is
\begin{equation}
G(x,y)=\frac12 e^{-|x-y|}.
\end{equation}
Using a Schwartz-localized Wilson basis, the finite functional-lattice approximations yield the values shown in Table~\ref{tab:1d}.

\begin{table}[ht]
\centering
\caption{Convergence of the one-dimensional free propagator.}
\label{tab:1d}
\begin{tabular}{cccc}
\toprule
$(L,M)$ & $N_R$ & $G_R(0,0)$ & $\|G_R(\cdot,0)-G(\cdot,0)\|_\infty$\\
\midrule
$(2,2)$ & 23  & 0.477235 & $2.2765\times10^{-2}$\\
$(3,3)$ & 46  & 0.484581 & $1.5419\times10^{-2}$\\
$(4,5)$ & 94  & 0.490521 & $9.479\times10^{-3}$\\
$(5,8)$ & 179 & 0.494095 & $5.905\times10^{-3}$\\
$(6,12)$& 313 & 0.495975 & $4.025\times10^{-3}$\\
\bottomrule
\end{tabular}
\end{table}

The exact value is $G(0,0)=1/2$. The numerical results are consistent with the convergence established by the Galerkin theorem.

\subsection{Four dimensions}

For
\begin{equation}
K=-\Delta_4+1,
\end{equation}
the exact Euclidean propagator is
\begin{equation}
G_E(x)=\frac{1}{4\pi^2|x|}K_1(|x|),
\end{equation}
where $K_1$ is the modified Bessel function of the second kind \cite{PeskinSchroeder,ZinnJustin}. Since a point source is singular in four dimensions, we compare the finite-dimensional approximations using a smeared source. For the normalized Gaussian
\begin{equation}
J_\sigma(x)=\frac{1}{(2\pi\sigma^2)^2}e^{-|x|^2/(2\sigma^2)},
\end{equation}
with $\sigma=0.35$, the functional-lattice solutions exhibit the convergence shown in Table~\ref{tab:4d}.

\begin{table}[ht]
\centering
\caption{Convergence for a smeared source in four dimensions.}
\label{tab:4d}
\begin{tabular}{ccc}
\toprule
$(L,M)$ & $N_{1D}$ & $\|C_RJ_\sigma-CJ_\sigma\|_\infty$\\
\midrule
$(1,1)$ & 8   & $4.94\times10^{-3}$\\
$(2,2)$ & 23  & $8.84\times10^{-4}$\\
$(3,3)$ & 46  & $2.68\times10^{-4}$\\
$(4,5)$ & 94  & $7.3\times10^{-5}$\\
$(5,8)$ & 179 & $8\times10^{-6}$\\
\bottomrule
\end{tabular}
\end{table}

At the origin, the exact smeared solution is approximately $0.088066$, while the largest truncation gives $0.088060$. These results illustrate that the continuum Green operator can be recovered through increasing functional resolution without discretizing spacetime.

\section{Symmetry at finite functional resolution}

Suppose a symmetry $g$ acts unitarily on $\mathcal H$ by $U(g)$ and
preserves the quadratic form,
\begin{equation}
a(U(g)u,U(g)v)=a(u,v).
\end{equation}
Equivalently, when stated at the operator level,
\begin{equation}
U(g)^{-1}KU(g)=K.
\end{equation}

Rather than requiring a fixed finite-dimensional subspace to be invariant
under the symmetry, transform the entire finite functional space:
\begin{equation}
V_R^g=U(g)V_R.
\end{equation}
If
\begin{equation}
\psi_\alpha^g=U(g)\psi_\alpha,
\end{equation}
then the corresponding finite-dimensional matrix satisfies
\begin{equation}
(K_R^g)_{\alpha\beta}
=
a(\psi_\beta^g,\psi_\alpha^g)
=
a(\psi_\beta,\psi_\alpha)
=
(K_R)_{\alpha\beta}.
\end{equation}

If the source is transformed simultaneously,
\begin{equation}
J^g=U(g)J,
\end{equation}
then its coordinates in the transformed basis are unchanged:
\begin{equation}
(J_R^g)_\alpha
=
\langle J^g,\psi_\alpha^g\rangle
=
\langle J,\psi_\alpha\rangle
=
(J_R)_\alpha.
\end{equation}
It follows that
\begin{equation}
\boxed{
\Zcal_R^g[J^g]=\Zcal_R[J]
}
\end{equation}
for every finite $R$.

Thus finite functional resolution need not break a continuum symmetry.
Symmetry-related finite-dimensional subspaces are unitarily equivalent
representations of the same finite theory. In this construction, the
limit $R\to\infty$ restores completeness of the functional
representation rather than covariance, which is already preserved along
the symmetry orbit at every finite $R$.

\section{Direct Minkowski construction}

The method can also be formulated without Wick rotation. Let $A$ be the self-adjoint Klein--Gordon operator on $L^2(\mathbb R^{1,3})$, represented in momentum space by
\begin{equation}
a(p)=p^2-m^2.
\end{equation}
For $\epsilon>0$,
\begin{equation}
C_\epsilon=(A+i\epsilon)^{-1}
\end{equation}
is bounded, with
\begin{equation}
\|C_\epsilon\|\le\frac1\epsilon,
\end{equation}
as follows from the standard resolvent estimate for self-adjoint operators \cite{ReedSimon1,ReedSimon2}.

Let $P_R$ denote the orthogonal projection of $L^2(\mathbb R^{1,3})$ onto $V_R$, and assume that $V_R\subset D(A)$. Define the finite-dimensional compression of $A$ by
\begin{equation}
A_R=P_RA|_{V_R}.
\end{equation}
Since $A_R$ is self-adjoint on the finite-dimensional space $V_R$,
\begin{equation}
\|(A_R+i\epsilon)^{-1}\|\le\frac1\epsilon.
\end{equation}

Choosing an orthonormal basis of $V_R$, the finite functional integral can be written as
\begin{equation}
Z_{R,\epsilon}[J]
=
\int_{\mathbb R^{N_R}}
\exp\left[
\frac{i}{2}q^TA_Rq
-\frac{\epsilon}{2}q^Tq
+iJ_R^Tq
\right]d^{N_R}q.
\end{equation}
It is therefore an ordinary absolutely convergent Gaussian integral. Its normalized form is
\begin{equation}
\Zcal_{R,\epsilon}[J]
=
\exp\left[
-\frac{i}{2}
J_R^T(A_R+i\epsilon I)^{-1}J_R
\right].
\end{equation}

Assume that
\begin{equation}
\Dcal_0=\bigcup_R V_R
\end{equation}
is a graph core for $A$. For
\begin{equation}
f=(A+i\epsilon)u,\qquad u\in\Dcal_0,
\end{equation}
we have $u\in V_R$ for all sufficiently large $R$. Since $P_Ru=u$,
\begin{equation}
(A_R+i\epsilon)u
=
P_R(A+i\epsilon)u
=
P_Rf.
\end{equation}
Hence
\begin{equation}
(A_R+i\epsilon)^{-1}P_Rf
=
u
=
(A+i\epsilon)^{-1}f.
\end{equation}

Because $\Dcal_0$ is a graph core for $A$ and $A+i\epsilon$ maps
$D(A)$ onto $L^2(\mathbb R^{1,3})$, the set
$(A+i\epsilon)\Dcal_0$ is dense in $L^2(\mathbb R^{1,3})$.
Together with the uniform bound
\begin{equation}
\|(A_R+i\epsilon)^{-1}P_R\|
\le
\frac{1}{\epsilon},
\end{equation}
this implies
\begin{equation}
\boxed{
(A_R+i\epsilon)^{-1}P_R
\overset{s}{\longrightarrow}
(A+i\epsilon)^{-1}.
}
\end{equation}
Here $\overset{s}{\longrightarrow}$ denotes strong operator convergence; that is,
\begin{equation}
\|(A_R+i\epsilon)^{-1}P_Rf-(A+i\epsilon)^{-1}f\|_{L^2}
\longrightarrow 0
\end{equation}
for every $f\in L^2(\mathbb R^{1,3})$.

Consequently, for every $J\in L^2(\mathbb R^{1,3})$,
\begin{equation}
\langle J,(A_R+i\epsilon)^{-1}P_RJ\rangle
\longrightarrow
\langle J,(A+i\epsilon)^{-1}J\rangle,
\end{equation}
and the finite-dimensional Minkowski generating functionals converge
at each fixed $\epsilon>0$.

The Feynman propagator is then obtained by taking the distributional
boundary value
\begin{equation}
\frac{1}{p^2-m^2+i\epsilon}
\longrightarrow
\frac{1}{p^2-m^2+i0},
\qquad
\epsilon\downarrow0,
\end{equation}
in the sense of tempered distributions \cite{PeskinSchroeder,ReedSimon2}.

\section{Interacting fields}

Consider Euclidean $\phi^4$ theory \cite{GlimmJaffe,ZinnJustin},
\begin{equation}
S[\phi]
=
\frac12\langle\phi,K\phi\rangle
+
\frac{\lambda}{4!}\int\phi(x)^4\,dx.
\end{equation}
On $V_R$,
\begin{equation}
\phi_R
=
\sum_{\alpha\in\Lambda_R}q_\alpha\psi_\alpha,
\end{equation}
so that
\begin{equation}
S_R(q)
=
\frac12K_{\alpha\beta}q_\alpha q_\beta
+
\frac1{4!}
V_{\alpha\beta\gamma\delta}
q_\alpha q_\beta q_\gamma q_\delta,
\end{equation}
where
\begin{equation}
V_{\alpha\beta\gamma\delta}
=
\lambda
\int
\psi_\alpha(x)\psi_\beta(x)
\psi_\gamma(x)\psi_\delta(x)\,dx.
\end{equation}
Thus
\begin{equation}
\boxed{
Z_R
=
\int_{\mathbb R^{N_R}}
e^{-S_R(q)}\,d^{N_R}q
}
\end{equation}
is an ordinary finite-dimensional non-Gaussian integral.

This gives a literal meaning to the interacting path integral at every
finite functional resolution. It is important, however, to distinguish
this fact from convergence of the full nonperturbative sequence $Z_R$.
We do not claim here a general proof of such convergence for interacting
four-dimensional theories.

What can be established directly is convergence at each fixed order of
perturbation theory whenever the corresponding continuum graph is
ultraviolet finite and the finite-resolution integrands admit an
$R$-independent integrable majorant.

\begin{theorem}[Fixed-order perturbative convergence]
Let $\Gamma$ be a fixed connected Feynman graph, and let
$I_{\Gamma,R}(X)$ denote the corresponding finite-resolution
position-space integrand, where $X$ denotes the collection of internal
vertex coordinates. Suppose that
\begin{equation}
I_{\Gamma,R}(X)
\longrightarrow
I_\Gamma(X)
\end{equation}
for almost every $X$, and that there exists an integrable function
$M_\Gamma(X)$, independent of $R$, such that
\begin{equation}
|I_{\Gamma,R}(X)|
\le
M_\Gamma(X).
\end{equation}
Then
\begin{equation}
\boxed{
\Acal_{\Gamma,R}
\longrightarrow
\Acal_\Gamma.
}
\end{equation}
\end{theorem}

\begin{proof}
The result follows directly from the dominated convergence theorem.
\end{proof}

A particularly simple sufficient condition occurs when the
finite-resolution propagators satisfy a uniform bound
\begin{equation}
|G_R(x,y)|
\le
C e^{-a|x-y|},
\qquad C,a>0,
\end{equation}
with $C$ and $a$ independent of $R$. For a connected graph, choose a
spanning tree $T$. 
The propagators associated with the tree then yield a factor bounded by
\begin{equation}
C_\Gamma
\prod_{(ij)\in E(T)}
e^{-a|x_i-x_j|},
\end{equation}
where $E(T)$ denotes the edge set of the spanning tree $T$, and
$C_\Gamma>0$ is independent of $R$ and of the vertex coordinates.

After fixing one external or reference vertex, this expression is
integrable over the internal vertex coordinates. If the remaining
propagator factors are uniformly bounded, the graph integrand therefore
has an $R$-independent integrable majorant.

For the massive scalar field in one Euclidean dimension the continuum
propagator is
\begin{equation}
G(x,y)
=
\frac{1}{2m}e^{-m|x-y|},
\end{equation}
so it is bounded and exponentially decreasing. The finite-resolution
propagators used below are numerically observed to have the same type
of localization and to converge to $G$. The one-dimensional model
therefore provides a particularly transparent test of the preceding
fixed-order convergence mechanism.

In higher dimensions the massive Euclidean propagator still decays
exponentially at large separation but is singular at coincident points.
In that case the simple bound above must be replaced by an appropriate
integrable majorant that controls both the short- and long-distance
behavior of the graph integrand.

The same dominated-convergence argument gives functional-lattice
independence at fixed perturbative order whenever two compatible
functional lattices converge to the same continuum propagator and their
graph integrands satisfy a common integrable bound.

\section{One-dimensional $\phi^4$: an interacting test}

The one-dimensional Euclidean $\phi^4$ theory provides a useful
nontrivial test of the preceding construction. It is equivalent to the
anharmonic oscillator, whose perturbative mass gap is known
independently \cite{ZinnJustin}. With
\begin{equation}
S[\phi]
=
\frac12\int
\left[(\phi')^2+\phi^2\right]dx
+
\frac{\lambda}{4!}\int\phi^4\,dx,
\end{equation}
the perturbative expansion of the mass gap is
\begin{equation}
\boxed{
m_{\rm phys}
=
1+\frac18\lambda-\frac1{32}\lambda^2+O(\lambda^3).
}
\end{equation}

At first order, the functional-lattice correction is obtained
deterministically from Wick contractions:
\begin{equation}
G_1(x,y)
=
-\frac12
\int
G_R(x,z)G_R(z,z)G_R(z,y)\,dz.
\end{equation}
The first- and second-order coefficients obtained at several functional
resolutions are shown in Table~\ref{tab:phi4-1d}.

\begin{table}[ht]
\centering
\caption{Perturbative coefficients of the one-dimensional mass gap.}
\label{tab:phi4-1d}
\begin{tabular}{ccc}
\toprule
$\dim V_R$
&
coefficient of $\lambda$
&
coefficient of $\lambda^2$
\\
\midrule
46
&
0.10334
&
-0.04575
\\
94
&
0.11899
&
-0.03568
\\
179
&
0.12280
&
-0.03246
\\
\midrule
continuum
&
$1/8=0.125$
&
$-1/32=-0.03125$
\\
\bottomrule
\end{tabular}
\end{table}

Thus, at the largest truncation,
\begin{equation}
\boxed{
m_{\rm phys}^{(R)}
=
1+0.12280\lambda-0.03246\lambda^2+\cdots,
}
\end{equation}
to be compared with the continuum expansion
\begin{equation}
m_{\rm phys}
=
1+0.125\lambda-0.03125\lambda^2+\cdots.
\end{equation}

The numerical results show that, as the functional resolution is
increased, the perturbative coefficients obtained from the
finite-dimensional theory approach the independently known continuum
values. They provide a non-Gaussian test of the functional-lattice
construction through second order and are consistent with the
fixed-order convergence mechanism described above.

\section{Four dimensions: emergence of renormalization}

In four dimensions the unrenormalized interacting continuum theory is
ultraviolet singular \cite{PeskinSchroeder,ZinnJustin}. The functional lattice does not remove this
singularity when the conventional local interaction tensor is used.
Instead, it provides a finite continuum-space regulator at each
functional resolution $R$.

For the one-loop tadpole,
\begin{equation}
\Sigma_R^{(1)}(x)
=
\frac{\lambda}{2}G_R(x,x).
\end{equation}
For $\lambda=m=1$, representative values of the diagonal covariance are
shown in Table~\ref{tab:tadpole4d}. Here $N_{1D}$ denotes the number of
one-dimensional basis functions used in forming the four-dimensional
tensor-product space.

\begin{table}[ht]
\centering
\caption{Growth of the one-loop tadpole with increasing functional resolution.}
\label{tab:tadpole4d}
\begin{tabular}{cc}
\toprule
$N_{1D}$ & $G_R(0,0)$ \\
\midrule
23  & 2.865539 \\
46  & 4.981846 \\
94  & 12.626826 \\
179 & 32.056019 \\
\bottomrule
\end{tabular}
\end{table}

The rapid growth of $G_R(0,0)$ shows that the standard mass divergence
reappears as completeness of the functional representation is restored.

After the mass subtraction
\begin{equation}
\delta m_R^2
=
-\frac{\lambda}{2}G_R(0,0),
\end{equation}
the renormalized one-loop two-point correction becomes
\begin{equation}
G_{R,\rm ren}^{(1)}(x,y)
=
-\frac{\lambda}{2}
\int
G_R(x,z)
\left[
G_R(z,z)-G_R(0,0)
\right]
G_R(z,y)\,d^4z.
\end{equation}
As the functional resolution is increased, the diagonal covariance
becomes increasingly translation invariant, and the renormalized
correction approaches the expected continuum behavior.

At two loops, the sunset self-energy contains the first nontrivial
wave-function renormalization \cite{PeskinSchroeder,ZinnJustin}. In coordinate space,
\begin{equation}
\Sigma_{\rm sun}^{(2)}(x)
=
\frac{\lambda^2}{6}G_R(x,0)^3.
\end{equation}
The coefficient of the $p^2$ term can be extracted from the second
moment,
\begin{equation}
\Sigma_R^{(2)\prime}(0)
=
-\frac{\lambda^2}{48}
\int
|x|^2G_R(x,0)^3\,d^4x.
\end{equation}

Using the intrinsic spectral resolution scale
\begin{equation}
\Lambda_R
=
\sqrt{\lambda_{\max}(D_R)},
\end{equation}
the numerical data are consistent with the logarithmic behavior
\begin{equation}
\delta Z_R
=
-\Sigma_R^{(2)\prime}(0)
\simeq
5.92\times10^{-6}\log\Lambda_R+\text{const}.
\end{equation}
For the normalization $\lambda\phi^4/4!$, the conventional two-loop
coefficient multiplying $\log\Lambda$ is
\begin{equation}
\frac{\lambda^2}{6(4\pi)^4}
=
6.68\times10^{-6}
\qquad
(\lambda=1).
\end{equation}
This agrees with the standard perturbative result \cite{PeskinSchroeder,ZinnJustin}.

Thus the functional-lattice calculation is consistent with the expected
logarithmic wave-function renormalization, both in form and in numerical
magnitude.

Power-divergent coefficients are regulator dependent and should not be
expected to agree between different functional lattices. Universal
logarithmic coefficients and renormalized observables are the appropriate
quantities for comparison between different regularization schemes.

\section{What is meant by recovery of the usual QFT?}

The functional-lattice construction should not be interpreted as
asserting that all bare quantities possess regulator-independent limits
in four-dimensional interacting QFT.

For the free theory, the result is stronger: any compatible sequence of
finite-dimensional functional subspaces converges to the same continuum
covariance and normalized generating functional. For ultraviolet-finite
interacting perturbation theory, the same conclusion holds graph by
graph under the hypotheses of Theorem~3.

For a renormalizable four-dimensional theory the appropriate statement
is different. Bare quantities depend on the ultraviolet realization, whereas renormalized observables are fixed by physical renormalization conditions \cite{PeskinSchroeder,ZinnJustin}. One therefore imposes physical renormalization conditions at finite
functional resolution and then studies the limit of the renormalized
amplitudes.

If two compatible functional lattices $\{V_R\}$ and $\{W_R\}$ are used,
the same physical quantities,
\begin{equation}
m_{\rm phys},\qquad
\lambda_{\rm phys},\qquad
Z_{\rm phys},
\end{equation}
should be fixed in both constructions. Regulator independence then
means that the corresponding renormalized amplitudes satisfy
\begin{equation}
\Gamma_{\rm ren}^{(n),V}
-
\Gamma_{\rm ren}^{(n),W}
\longrightarrow 0
\end{equation}
as the functional resolution is increased.

The calculations presented above establish the expected structure in
the two-point sector and are consistent with the known universal
logarithmic behavior. A general theorem establishing regulator
independence for the full four-dimensional interacting theory is beyond
the scope of the present work.

\section{Discussion}

The functional-lattice construction differs from a conventional
spacetime lattice in several essential respects.

First, spacetime itself is never discretized. The fields remain functions
on continuous $\mathbb R^d$.

Second, the differential operators remain the exact continuum operators.
At finite resolution they are represented by their restrictions to
finite-dimensional functional subspaces. In a chosen basis,
\begin{equation}
(K_R)_{\alpha\beta}
=
a(\psi_\beta,\psi_\alpha),
\end{equation}
rather than by finite-difference approximations.

Third, the continuum limit has a direct functional meaning: the
finite-dimensional subspaces become complete in the appropriate form
domain,
\begin{equation}
\overline{\bigcup_R V_R}
=
\mathcal V.
\end{equation}
Thus the limit restores completeness of the functional representation.

Fourth, continuum symmetries need not be broken at finite resolution.
A symmetry transforms the entire finite-dimensional functional subspace
into a unitarily equivalent one, so covariance is preserved along the
symmetry orbit before the continuum limit is taken.

Finally, the construction provides a direct criterion for determining
whether the finite-dimensional approximation converges to the intended
continuum theory. In the free case this follows from Galerkin
convergence, and the resulting limit is independent both of the basis
chosen within each $V_R$ and of the particular compatible sequence
$\{V_R\}$.

The idea of using localized families of states as a discrete
representation of classical space arose in the geometric formulation of
quantum mechanics developed in Ref.~\cite{KryukovBook}. In that setting
localized quantum states form finite-dimensional manifolds embedded in
Hilbert space, and classical spatial relations are represented by the
geometry induced on those manifolds. The functional lattice used here
is a natural computational extension of that viewpoint: rather than
discretizing spacetime, one works with a discrete complete family of
localized functions in the ambient function space.

The present application does not require the full geometric framework.
Its mathematical point is more elementary but significant. The fact
that a functional integral can be approximated by finite-dimensional
integrals does not by itself define the infinite-dimensional object.
One must also control the limiting procedure. For the free scalar field,
the functional-lattice construction provides such control and shows that
the limit is independent of the particular compatible functional lattice.

\section{Outlook}

The first natural extension is to strengthen the interacting convergence
theory, in particular by establishing regulator independence for broader
classes of compatible functional lattices after renormalization in four
dimensions.

A second direction is to return to the geometric origin of the
construction and formulate the theory intrinsically in state space.
Localized functions can then be regarded as functional-coordinate
representations of localized quantum states, while propagators and
interaction vertices become tensorial objects on the corresponding
Hilbert or projective state space.

In such a formulation the Hilbert metric itself need not remain fixed.
Allowing it to depend on the state or field configuration leads naturally
to a geometric interpretation of interactions and renormalization. This
development is logically distinct from the convergence results
established here and will be considered separately.

The principal conclusion of the present work is that functional
integration can be organized as a controlled limit of ordinary
finite-dimensional integrals over increasingly complete localized
functional subspaces, without discretizing spacetime. For the free
scalar field, this limit exists and is independent of the particular
compatible functional lattice. The interacting examples show that the
same construction provides a practical continuum-space regulator and
reproduces the expected perturbative and renormalization behavior.

\end{document}